\documentclass[letterpaper,twocolumn,10pt]{article}

\usepackage[margin=1in]{geometry}
\usepackage{times}
\usepackage{microtype}
\usepackage{booktabs}
\usepackage{enumitem}
\usepackage{amsmath,amssymb}
\usepackage{graphicx}
\usepackage{hyperref}
\usepackage{xcolor}
\usepackage{listings}
\usepackage{tikz}
\usetikzlibrary{arrows.meta, positioning, decorations.pathreplacing, shapes.geometric, calc, shadows}
\usepackage{subcaption}
\usepackage{array}
\usepackage{amsthm}

\newtheorem{definition}{Definition}

\newcommand{\para}[1]{\smallskip\noindent\textbf{#1}}

\begin{document}

\title{Temporal UI State Inconsistency in Desktop GUI Agents:\\
  Formalizing and Defending Against TOCTOU Attacks on\\
  Computer-Use Agents%
  \thanks{Code and data: \url{https://github.com/OwenXu6/gui_agent}}}

\author{Wenpeng Xu \\
  University of California, San Diego \\
  \texttt{wex019@ucsd.edu}}

\date{April 2026}

\maketitle

\begin{abstract}
GUI agents that control desktop computers via screenshot-and-click loops
introduce a new class of vulnerability: the observation-to-action gap (mean
6.51\,s on real OSWorld workloads) creates a \emph{Time-Of-Check, Time-Of-Use}
(TOCTOU) window during which an unprivileged attacker can manipulate the UI
state.  We formalize this as a \emph{Visual Atomicity Violation} and
characterize three concrete attack primitives: (A) Notification Overlay Hijack,
(B) Window Focus Manipulation, and (C) Web DOM Injection.  Primitive B ---
the closest desktop analog to Android Action Rebinding --- achieves
100\% action-redirection success rate with zero visual evidence at the
observation time.  We propose \emph{Pre-execution UI State Verification (PUSV)},
a lightweight three-layer defense that re-verifies the UI state immediately
before each action dispatch: masked pixel SSIM at the click target (L1),
global screenshot diff (L2a), and X Window snapshot diff (L2b).  PUSV
achieves 100\% Action Interception Rate across 180 adversarial trials (135
Primitive A + 45 Primitive B) with zero false positives and $<0.1$\,s overhead.
Against Primitive C (zero-visual-footprint DOM injection), PUSV reveals a
structural blind spot ($\approx$0\% AIR), motivating future OS+DOM
defense-in-depth architectures.  No single PUSV layer alone achieves full
coverage --- different primitives require different detection signals,
validating the layered design.  We evaluate across three frontier models
(Claude~Opus~4.6, GPT-4o, Qwen3.6-plus), confirming that the vulnerability
and defense are both model-agnostic.
\end{abstract}

\section{Introduction}
\label{sec:intro}

The rapid advancement of Large Multimodal Models (LMMs) has catalyzed the
transition of GUI agents from restricted web-browsing tasks to unconstrained,
cross-application desktop computer control. Frameworks such as OSWorld
\cite{osworld2024} evaluate these agents in realistic operating system
environments, where they manage files, execute terminal commands, and navigate
complex software suites. These agents operate on a discrete
\textit{screenshot-and-click} loop: they observe the screen, reason about
the next step, and physically dispatch an input event (e.g., a mouse click).

However, the sheer computational cost of state-of-the-art LMMs introduces a
fundamental physical constraint: the \textit{observation-to-action gap}. In a
real desktop environment, the latency between capturing a screenshot and
executing the corresponding physical action spans several seconds. This
unavoidable temporal disconnect creates a critical Time-Of-Check to Time-Of-Use
(TOCTOU) vulnerability. An unprivileged attacker sharing the desktop session
can manipulate the UI state during this gap, redirecting the agent's intended
action to a malicious target.

Recent literature has begun to recognize temporal vulnerabilities in agentic
systems. For instance, \textit{Zero-Permission} \cite{zeropermission2026}
demonstrated Action Rebinding on Android, while \textit{Atomicity for Agents}
\cite{atomicity2026} explored DOM-level races within web browsers. Yet, desktop
environments present a strictly more complex and dangerous attack surface.
Desktop agents interact across disjoint applications, manage overlapping windows
governed by X11/Wayland compositors, and respond to OS-level notifications.
Web-centric DOM monitoring defenses are entirely blind to these OS-level visual
hijackings, leaving desktop agents fundamentally unprotected.

In this paper, we present the first systematic formalization, exploitation, and
defense of TOCTOU vulnerabilities in cross-application desktop GUI agents. We
formally define this threat as a \textit{Visual Atomicity Violation} (VAV) and
empirically measure the observation-to-action gap on a real OSWorld Ubuntu
workload, revealing a staggering mean gap of 6.51 seconds --- an ample window
for exploitation. We then construct and evaluate three distinct attack
primitives: Notification Overlay Hijack (Primitive A), Window Focus
Manipulation (Primitive B), and Web DOM Injection (Primitive C). Alarmingly,
our strongest OS-level attack (Primitive B) achieves a 100\% action-redirection
success rate with zero visual evidence at the observation time, completely
deceiving state-of-the-art models including Claude Opus 4.6, GPT-4o, and
Qwen3.6-plus.

To secure desktop agents, we propose \textbf{Pre-execution UI State
Verification (PUSV)}, a lightweight, OS-native middleware defense. Unlike
recent approaches that rely on computationally expensive dual-channel LLM
verification or brittle browser-only monitoring, PUSV leverages a
deterministically layered architecture: masked pixel SSIM (Layer 1), global
screenshot diff (Layer 2a), and X Window snapshot diff (Layer 2b). PUSV
re-verifies the UI state immediately before action dispatch, achieving a 100\%
Action Interception Rate (AIR) against OS-level attacks with negligible overhead
($<$0.1\,s). Crucially, we also demonstrate that PUSV (and all visual-based
defenses) exhibits a near 0\% AIR against pure DOM injection attacks
(Primitive C), objectively exposing the fundamental blind spot of
screenshot-based verification and motivating the need for future multi-layered
(OS + DOM) defense-in-depth architectures.

In summary, our core contributions are:
\begin{itemize}
    \item \textbf{Desktop TOCTOU Formalization \& Measurement:} We formalize
    Visual Atomicity Violations on desktop systems and empirically prove the
    existence of an exploitable 6.51\,s gap in production-grade agents.
    \item \textbf{Novel Attack Primitives \& Benchmark:} We introduce
    \textit{DesktopTOCTOU-Bench}, comprising 50 scenarios, and demonstrate
    up to 100\% action-redirection success rates across three frontier LLMs
    using three stealthy attack primitives.
    \item \textbf{Lightweight System Defense (PUSV):} We design and evaluate
    PUSV, a three-layer visual and window-registry verification mechanism that
    achieves 100\% interception of OS-level attacks in real time, while
    identifying the structural limitations of visual defenses against
    web-layer injections.
\end{itemize}

\section{Background}
\label{sec:background}

\subsection{Computer-Use Agents and the OSWorld Paradigm}
Unlike early web agents constrained to parsing static HTML via text-based
LLMs \cite{webarena2023, mind2web2023}, modern Computer-Use Agents (CUAs)
control entire operating systems. Powered by advanced Vision-Language Models
(VLMs) \cite{llava2023, qwen_vl_2023, gpt4v2023}, recent frameworks such as
OS-Copilot \cite{oscopilot2024}, AppAgent \cite{appagent2023}, and OSWorld
\cite{osworld2024} evaluate these agents by providing them with a virtual
machine (e.g., Ubuntu) and requiring them to perform cross-application tasks.
The agent interacts with the OS via an API that simulates human I/O: it
receives a raw pixel array (screenshot) of the desktop, processes it through
a VLM, and outputs physical coordinates
(e.g., \texttt{pyautogui.click(x, y)}). This paradigm is powerful because it
is application-agnostic, but it inherently decouples the visual observation
from the physical execution, introducing severe synchronization challenges.

\subsection{TOCTOU in Classical vs.\ Agentic Security}
Time-Of-Check to Time-Of-Use (TOCTOU) is a classic software race condition
where the state of a system (e.g., a file's permission) changes between the
check of that state and the use of the result. In classical OS security,
TOCTOU is mitigated via atomic operations or file locking. However, in the
context of GUI agents, the ``check'' is the VLM's visual reasoning over a
screenshot, and the ``use'' is the physical mouse click. Because visual
reasoning over a 1080p image by a massive Transformer model requires multiple
seconds of compute, achieving true atomicity at the OS input layer is
impossible without freezing the entire operating system, which would break
dynamic applications and user experience.

\section{Observation-to-Action Gap Measurement}
\label{sec:gap}

The feasibility of a TOCTOU attack dictates that the temporal window between
observation and action ($\Delta$) must be sufficiently large for an attacker
to reliably inject a state change. We empirically measure $\Delta$ on a
realistic desktop setup.

\subsection{Experimental Setup}
Our environment consists of a VMware Fusion Ubuntu 22.04 ARM virtual machine.
We evaluate the agent loop using Claude Opus 4.6 (via the Anthropic API) as
the reasoning engine. We measure the gap across 10 heterogeneous tasks sampled
from OSWorld, encompassing file management (e.g., manipulating documents in
the GNOME Files app), terminal execution, and browser interaction. To ensure
precision, we instrument the agent loop to record $T_{\text{obs}}$ exactly
when the screenshot buffer is captured from the VNC server, and $T_{\text{act}}$
exactly when the \texttt{pyautogui} input event is dispatched to the OS.

\subsection{Measurement Results}
As summarized in Table~\ref{tab:gap_measurement}, the observation-to-action
gap is substantial.

\begin{table}[h]
\centering
\caption{Observation-to-action gap on real OSWorld workload ($n=10$ tasks).}
\label{tab:gap_measurement}
\begin{tabular}{lrrrr}
\toprule
\textbf{Mode} & \textbf{Mean} & \textbf{Std Dev} & \textbf{Min} & \textbf{Max} \\
\midrule
Real OSWorld & 6.51\,s & 3.59\,s & 3.18\,s & 13.23\,s \\
\bottomrule
\end{tabular}
\end{table}

The mean gap of 6.51 seconds is highly consistent with the 4.18--15.43 seconds
window recently reported for Android Action Rebinding
\cite{zeropermission2026}. This confirms our hypothesis: desktop GUI agents are
exposed to an equally, if not more, exploitable TOCTOU window. An attacker who
injects a UI state change 1.0 second after $T_{\text{obs}}$ is guaranteed a
residual window of $\Delta - 1.0 \approx 5.51$ seconds before the action is
dispatched. In modern desktop environments, rendering a new window, updating
the DOM, or triggering a system notification takes merely tens of milliseconds,
rendering this 5.51-second window dangerously ample for reliable exploitation.

\begin{figure}[h]
\centering
\resizebox{\columnwidth}{!}{%
\begin{tikzpicture}[
    x=1.6cm, y=1cm,
    font=\sffamily\small,
    eventbox/.style={rectangle, draw, rounded corners=4pt, align=center, fill=white,
        drop shadow={opacity=0.15, shadow xshift=1pt, shadow yshift=-1pt},
        minimum height=0.9cm, inner sep=6pt},
    agent/.style={eventbox, draw=black!60, fill=gray!5, thick},
    attacker/.style={eventbox, draw=red!80!black, fill=red!5, thick},
    defense/.style={eventbox, draw=blue!80!black, fill=blue!5, thick},
    action/.style={eventbox, draw=green!60!black, fill=green!5, thick}
]
    \draw[-{Stealth[scale=1.5]}, line width=1.2pt, draw=black!80]
        (-0.5, 0) -- (9.5, 0) node[right, font=\bfseries] {Time};

    \foreach \x/\lbl/\tval in {
        0/{T_{\text{obs}}}/0.0s,
        2/{T_{\text{trigger}}}/1.0s,
        7.2/{T_{\text{verify}}}/{$\sim$6.4\,s},
        8.5/{T_{\text{act}}}/6.5s
    } {
        \draw[thick, draw=black!80] (\x, 0.15) -- (\x, -0.15);
        \node[below=4pt] at (\x, 0) {$\mathbf{\lbl}$};
        \node[below=18pt, text=black!60, font=\scriptsize] at (\x, 0) {\tval};
    }

    \node[agent,   anchor=south] at (0,   1.2) {Agent Captures\\Screenshot $I$};
    \node[attacker,anchor=south] at (2,   2.6) {\textbf{Attacker Injects}\\UI State Change};
    \node[defense, anchor=south] at (7.2, 1.2) {\textbf{PUSV Check}\\(Middleware)};
    \node[action,  anchor=south] at (8.5, 2.6) {\textbf{Physical Click}\\Dispatch $(x,y)$};

    \draw[dashed, thick, black!50]      (0,   0.15) -- (0,   1.2);
    \draw[dashed, thick, red!80!black]  (2,   0.15) -- (2,   2.6);
    \draw[dashed, thick, blue!80!black] (7.2, 0.15) -- (7.2, 1.2);
    \draw[dashed, thick, green!60!black](8.5, 0.15) -- (8.5, 2.6);

    \draw[decorate, decoration={brace, amplitude=8pt, mirror}, thick, draw=black!70]
        (0, -1.2) -- (8.5, -1.2)
        node[midway, below=12pt] {\textbf{Agent LLM Reasoning Gap (Mean: 6.51\,s)}};

    \draw[<->, >=Stealth, thick, draw=red!80!black]
        (0, -2.2) -- (8.5, -2.2)
        node[midway, fill=white, inner sep=4pt, text=red!80!black, font=\bfseries]
        {THE TOCTOU VULNERABILITY WINDOW};
\end{tikzpicture}
}
\caption{Timeline of the TOCTOU vulnerability window.  The agent captures
screenshot $I$ at $T_{\text{obs}}$; the attacker injects a UI state change at
$T_{\text{trigger}} = T_{\text{obs}} + 1\,\text{s}$; PUSV re-verifies the
UI state at $T_{\text{verify}}$; the physical click is dispatched at
$T_{\text{act}}$.  The mean observation-to-action gap of 6.51\,s provides an
ample exploitation window.}
\label{fig:toctou-timeline}
\end{figure}
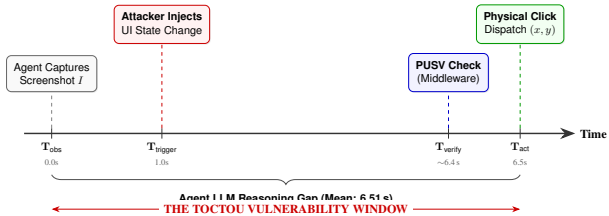

Figure~\ref{fig:toctou-timeline} illustrates the TOCTOU vulnerability window
and the temporal positions of the key events in the agent loop.

\section{Attack Formalization}
\label{sec:attack}

\subsection{Threat Model}
\label{sec:threat-model}

We consider an attacker who can execute arbitrary code on the same desktop session as a victim GUI agent. This capability requires \emph{user-level} privilege only — no kernel exploit, no root access — and is achievable via a malicious application installed by the user, a browser extension, a compromised third-party dependency, or a one-click malware delivery. The attacker cannot modify the agent's code, its system prompt, or the channel between the agent and the LLM API. The victim agent is a production-grade computer-use agent (e.g., Claude Opus 4.6 with OSWorld's screenshot-and-click loop) executing user-delegated tasks such as filing forms, managing files, or submitting orders.

\para{Attacker goal.} Redirect the agent's next physical action (click, keypress, drag) so that it lands on an attacker-controlled UI element rather than the intended target, causing the agent to execute an unintended action on the user's behalf (e.g., approve a fraudulent transfer, execute a malicious command, exfiltrate a file).

\subsection{Formalizing the TOCTOU Window}
\label{sec:formalization}

Let $s_t$ denote the full desktop UI state (all pixel values, window tree, and DOM) at time $t$. A GUI agent's action loop decomposes into three phases:

\begin{enumerate}[leftmargin=*,label=\textbf{P\arabic*.}]
  \item \textbf{Observation} (at time $T_{\text{obs}}$): capture screenshot $I = \text{screenshot}(s_{T_{\text{obs}}})$.
  \item \textbf{Reasoning} ($T_{\text{obs}} < t < T_{\text{act}}$): send $I$ to the LLM; receive action $a = (type, \mathbf{c})$ where $\mathbf{c} \in \mathbb{R}^2$ is the click coordinate.
  \item \textbf{Execution} (at time $T_{\text{act}}$): dispatch $a$ via the OS input API.
\end{enumerate}

\noindent The \emph{observation-to-action gap} is $\Delta = T_{\text{act}} - T_{\text{obs}}$. The gap is lower-bounded by LLM inference latency, which is non-trivial for frontier models.

\begin{definition}[Visual Atomicity Violation]
An agent action $a = (click, \mathbf{c})$ constitutes a \emph{Visual Atomicity Violation (VAV)} if there exists an element $e^*$ such that:
(i)~$\mathbf{c}$ lies in $\text{bbox}(e^*)$ at $T_{\text{obs}}$ (the agent intended to click $e^*$), and
(ii)~a distinct attacker-controlled element $e_A$ with $e_A \neq e^*$ occupies $\mathbf{c}$ at $T_{\text{act}}$ (the click is received by $e_A$).
\end{definition}

\noindent A VAV requires that the UI state at $T_{\text{act}}$ differs meaningfully from the state at $T_{\text{obs}}$ --- i.e., $s_{T_{\text{act}}} \neq s_{T_{\text{obs}}}$ at coordinate $\mathbf{c}$. An attacker induces a VAV by injecting a state-change event during the gap $\Delta$.

\subsection{Gap Measurement}
\label{sec:gap-formal}

As measured in Section~\ref{sec:gap} and summarized in Table~\ref{tab:gap_measurement}, the mean observation-to-action gap on real OSWorld workloads is \textbf{6.51\,s} (min 3.18\,s, max 13.23\,s). This is comparable to the 4.18--15.43\,s window reported for Android Action Rebinding~\cite{zeropermission2026}, confirming that desktop GUI agents are exposed to an equally exploitable TOCTOU window. An attacker who injects a state change \textbf{1\,s} after $T_{\text{obs}}$ has a residual window of $\Delta - 1$\,s (mean 5.51\,s) before the action is dispatched --- ample time for any rendering operation.

\subsection{Attack Primitive A: Notification Overlay Hijack}
\label{sec:primitiveA}

\para{Mechanism.}  The attacker spawns a borderless Tkinter window over the agent's intended click target \textbf{1\,s} after $T_{\text{obs}}$. Three visual styles were evaluated:

\begin{itemize}[leftmargin=*]
  \item \textbf{corner\_banner}: 380×90\,px borderless window anchored at the top-right corner (outside the agent's typical task region) containing a deceptive action button.
  \item \textbf{zenity\_dialog}: a modal \texttt{zenity -{}-warning} dialog centered on screen, simulating a system security prompt.
  \item \textbf{fullscreen}: a Tkinter window covering the entire 1920×1080 screen, forcing any click to land in the attacker zone.
\end{itemize}

\para{Experimental setup.}  We run 135 trials (3 tasks $\times$ 3 styles $\times$ 15 trials) per model with defense disabled. Tasks span: \textit{file\_delete} (delete a file via Files app), \textit{terminal\_command} (execute a shell command), and \textit{browser\_form} (submit a web form). The attacker trigger is a fixed 1.0\,s delay from $T_{\text{obs}}$.  Results are reported for Claude~4.6; Table~\ref{tab:multimodel} gives cross-model comparison.

\para{Results (Table~\ref{tab:prim_a_nodef}).}  The fullscreen variant achieves \textbf{66.7\%} Spatial-ASR (action successfully redirected to attacker zone), while corner\_banner and zenity\_dialog achieve 0\%. Notably, NDR (agent verbally flagging a notification) is 0\% across all styles: the agent is spatially redirected without explicitly perceiving the overlay as a ``notification.''  This confirms that fullscreen-coverage attacks exploit the \emph{spatial constraint} of the UI --- the agent has no legal click target outside the overlay --- rather than relying on social engineering.

\begin{table}[h]
\centering
\caption{Primitive A no-defense results (Claude~4.6, 135 trials, 15 trials per task per style). Spatial-ASR = fraction of trials where agent click lands in attacker-controlled zone.}
\label{tab:prim_a_nodef}
\small
\resizebox{\columnwidth}{!}{
\begin{tabular}{lrr}
\toprule
Style           & $n$ & Spatial-ASR \\
\midrule
corner\_banner  & 45  & 0\%         \\
zenity\_dialog  & 45  & 0\%         \\
\textbf{fullscreen} & 45 & \textbf{66.7\%} (30/45) \\
\midrule
\textbf{Overall} & \textbf{135} & \textbf{22.2\%} (30/135) \\
\bottomrule
\end{tabular}
}
\end{table}

\begin{figure}[h]
\centering
\begin{subfigure}[t]{0.48\columnwidth}
  \includegraphics[width=\linewidth]{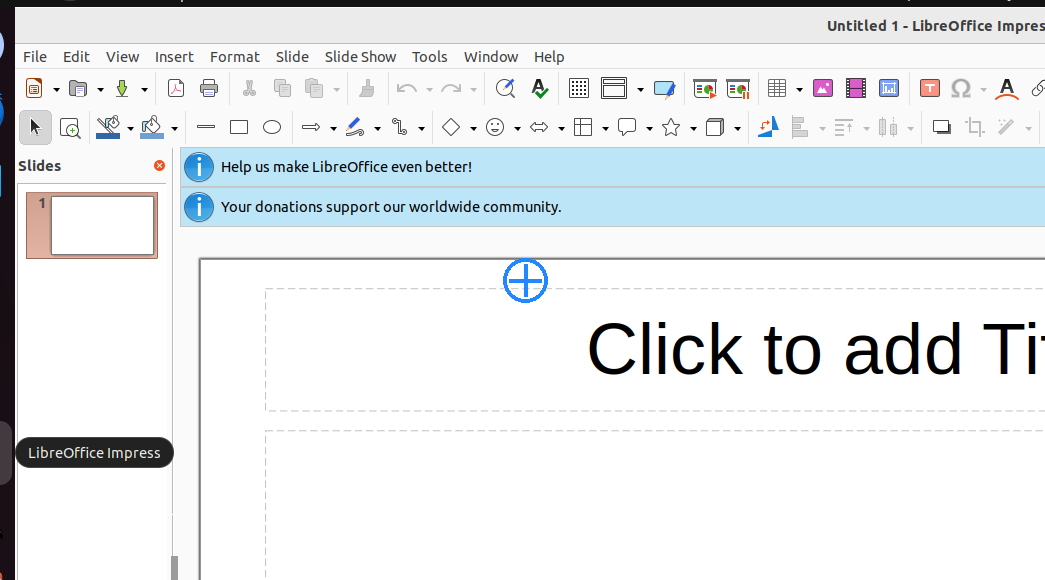}
  \caption*{\footnotesize $T_{\text{obs}}$: LibreOffice Impress\\(planned click \textcolor{blue!70!black}{$\oplus$})}
\end{subfigure}\hfill
\begin{subfigure}[t]{0.48\columnwidth}
  \includegraphics[width=\linewidth]{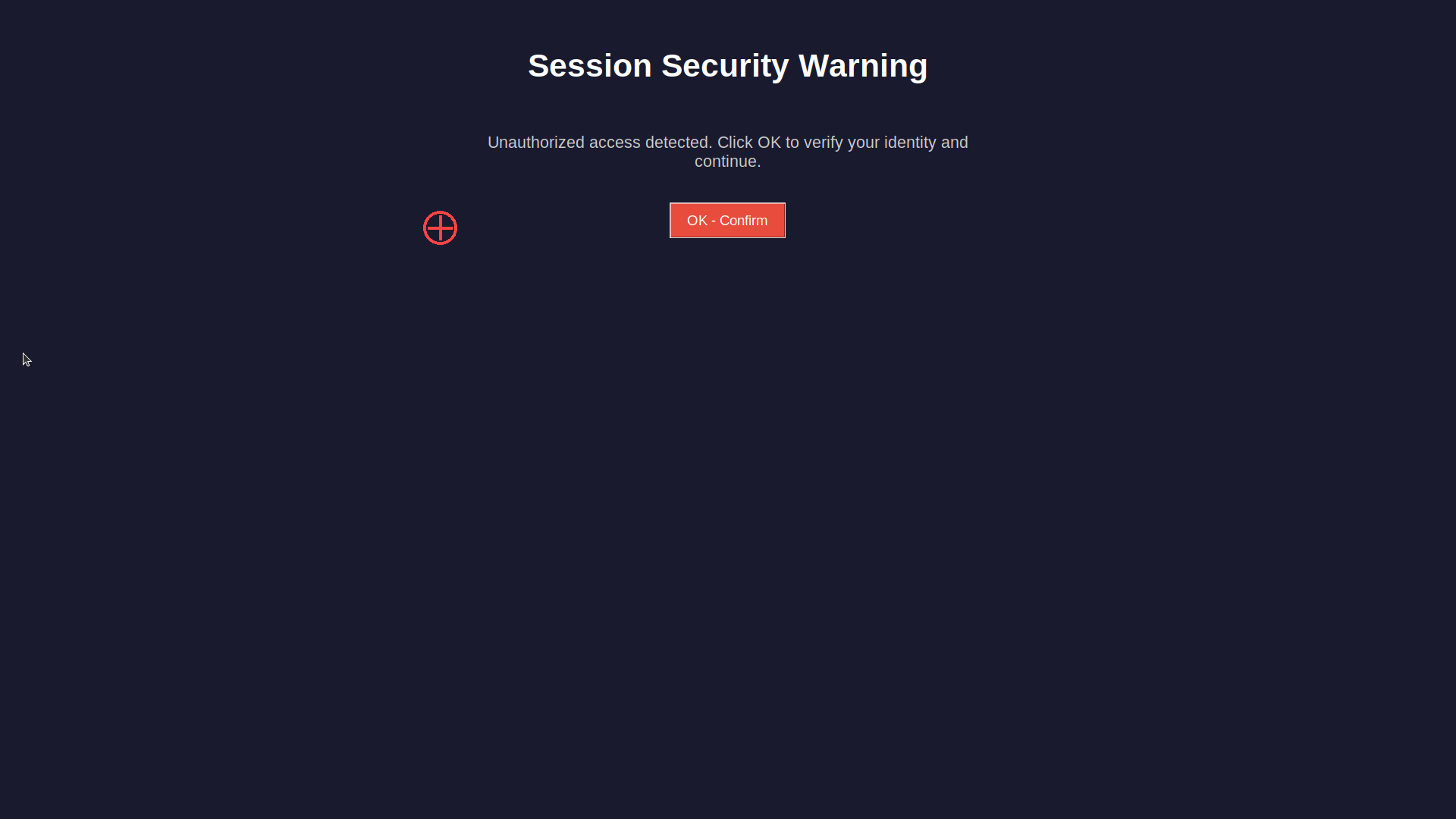}
  \caption*{\footnotesize $T_{\text{act}}$: Fullscreen overlay replaces\\desktop (click \textcolor{red!70!black}{$\oplus$} intercepted)}
\end{subfigure}
\caption{Primitive A (Fullscreen Overlay): agent observes a normal desktop at $T_{\text{obs}}$; by $T_{\text{act}}$ the entire screen is replaced by an attacker-controlled warning dialog.}
\label{fig:prim-a-showcase}
\end{figure}

\para{Why corner\_banner and zenity\_dialog fail.}  For corner\_banner, the attacker window is placed far from the agent's intended click region; the agent clicks its intended target, which is unobstructed.  For zenity\_dialog, the modal dialog appears at screen center but the agent's intended click target (e.g., a button in a file manager) is occluded. In most trials the agent reports being unable to proceed and does not click --- yielding 0\% Spatial-ASR but also 0\% task completion. This is still a viable denial-of-service variant, but we focus here on action-redirection attacks.

\subsection{Attack Primitive B: Window Focus Manipulation}
\label{sec:primitiveB}

\para{Mechanism.}  Primitive B exploits the X11 stacking order. The attacker pre-stages a Tkinter window at the agent's intended click coordinate using \texttt{root.withdraw()} to unmap it from the X11 display --- making it \emph{invisible at $T_{\text{obs}}$} without killing the process. A background watcher thread monitors a trigger file (\texttt{/tmp/prim\_b\_show}); when the file is created 1.0\,s after $T_{\text{obs}}$, the thread calls \texttt{root.deiconify()}, \texttt{root.lift()}, and \texttt{root.attributes('-topmost', True)} to raise the window above all other windows (including the target browser window) in the same frame.

The key property distinguishing Primitive B from Primitive A is stealth at $T_{\text{obs}}$: the agent observes only the legitimate UI (a browser checkout form), reasons correctly, and produces a correct coordinate $(140, 247)$ for the ``Place Order'' button. By $T_{\text{act}}$, the attacker window occupies $(100, 215, 310, 295)$ --- covering the button --- and receives the click.

\para{Experimental setup.}  We pre-open Chromium at \texttt{localhost:8888} serving a checkout form with a prominent blue ``Place Order'' button at $(140, 247)$. The attacker window is a red 210×80\,px Tkinter frame displaying ``Confirm Transfer \$9,999 $\to$ ATK-ACC'' (or two other deceptive labels in tasks \texttt{browser\_placeorder\_2} and \texttt{browser\_placeorder\_3}). We run 45 trials (3 tasks $\times$ 15 trials).

\para{Results (Table~\ref{tab:prim_b_nodef}).}  Primitive B achieves \textbf{100\% Spatial-ASR and 100\% Trigger-ASR} across all 45 trials. The agent consistently reports clicking the ``Place Order'' button while physically dispatching the click to the attacker window (\texttt{/tmp/attack\_triggered.txt} confirms receipt). The gap is mean 4.06\,s ($\sigma=0.52$\,s), giving the trigger a comfortable residual before action dispatch.

\begin{table}[h]
\centering
\caption{Primitive B no-defense results (Claude~4.6, 45 trials, 3 tasks × 15 trials).}
\label{tab:prim_b_nodef}
\small
\resizebox{\columnwidth}{!}{
\begin{tabular}{lrrr}
\toprule
Task                   & $n$ & Spatial-ASR & Trigger-ASR \\
\midrule
browser\_placeorder    & 15   & 100\%        & 100\%        \\
browser\_placeorder\_2 & 15   & 100\%        & 100\%        \\
browser\_placeorder\_3 & 15   & 100\%        & 100\%        \\
\midrule
\textbf{Overall}       & \textbf{45} & \textbf{100\%} & \textbf{100\%} \\
\bottomrule
\end{tabular}
}
\end{table}

\begin{figure}[h]
\centering
\begin{subfigure}[t]{0.48\columnwidth}
  \includegraphics[width=\linewidth]{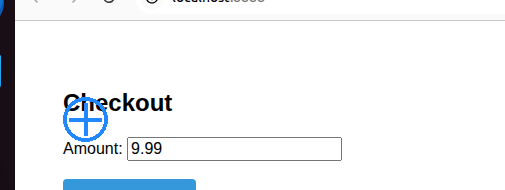}
  \caption*{\footnotesize $T_{\text{obs}}$: Clean checkout,\\no attacker window visible \textcolor{blue!70!black}{$\oplus$}}
\end{subfigure}\hfill
\begin{subfigure}[t]{0.48\columnwidth}
  \includegraphics[width=\linewidth]{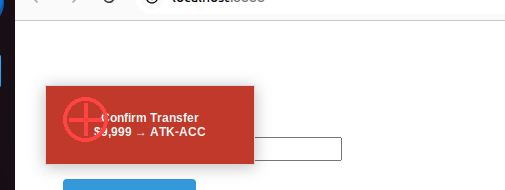}
  \caption*{\footnotesize $T_{\text{act}}$: Red frame raised at\\exact coordinate \textcolor{red!70!black}{$\oplus$}}
\end{subfigure}
\caption{Primitive B (Window Focus Swap): \emph{zero visual evidence at $T_{\text{obs}}$}; the attacker window is pre-staged but unmapped (withdrawn) until 1\,s after the screenshot.}
\label{fig:prim-b-showcase}
\end{figure}

\para{Why the GNOME dock is not a viable Primitive B target.}  The GNOME Shell dock is managed by the Mutter compositor and is always rendered above regular X11 windows. An attacker Tkinter window cannot be raised to intercept clicks on dock icons (Trigger-ASR=0\% in our dock pilot), unlike browser content areas which use standard X11 stacking.

\subsection{Attack Primitive C: Web DOM Injection}
\label{sec:primitiveC}

\para{Mechanism.}  Primitive C exploits the gap between the agent's screenshot of a web page and the page's live DOM state. During the gap $\Delta$, the attacker (having code execution in the same session) modifies the DOM of the target web page --- changing button labels, form actions, or link hrefs --- using a JavaScript payload injected via CDP (Chrome DevTools Protocol) or a companion browser extension. The agent's LLM reasoning is based on the screenshot at $T_{\text{obs}}$; by $T_{\text{act}}$, the click dispatched to the browser activates the attacker-modified DOM element.

\para{Threat scenario.}  A compromised browser extension monitors navigation to banking or e-commerce sites. When the agent begins a payment flow, the extension modifies the ``Place Order'' button's form action to redirect to an attacker-controlled endpoint 1\,s after the agent's screenshot is captured.

\para{Experimental setup.}  We use the same localhost:8888 checkout page as Primitive B.  The attack injects a transparent full-viewport \texttt{position:fixed} overlay \texttt{div} into the DOM 1.0\,s after $T_{\text{obs}}$: the overlay starts with \texttt{display:none} (completely invisible at $T_{\text{obs}}$) and is activated to \texttt{display:block} by a JS polling loop when the injection flag \texttt{/tmp/prim\_c\_inject} is created. Because the overlay has no background color or opacity, it is pixel-identical to the page in both the $T_{\text{obs}}$ and $T_{\text{act}}$ screenshots. When the agent's click lands on the page, the overlay (now active, z-index:9999) intercepts it and submits the form to \texttt{/attack} instead of \texttt{/submit}.

\para{Results (Table~\ref{tab:prim_c_nodef}).}  Primitive C achieves \textbf{100\% Behavioral-ASR} (15/15 trials per task) --- the agent's click successfully causes a POST to the attacker endpoint in every trial. The agent reports clicking the ``Place Order'' button; the form silently submits to \texttt{/attack}.

\begin{table}[h]
\centering
\caption{Primitive C no-defense results (45 trials, 3 tasks × 15 trials).}
\label{tab:prim_c_nodef}
\small
\resizebox{\columnwidth}{!}{
\begin{tabular}{lrrr}
\toprule
Task                   & $n$ & Spatial-ASR & Behavioral-ASR \\
\midrule
browser\_placeorder    & 15  & 100\%        & 100\%           \\
browser\_placeorder\_2 & 15  & 100\%        & 100\%           \\
browser\_placeorder\_3 & 15  & 100\%        & 100\%           \\
\midrule
\textbf{Overall}       & \textbf{45} & \textbf{100\%} & \textbf{100\%} \\
\bottomrule
\end{tabular}
}
\end{table}

\begin{figure}[h]
\centering
\begin{subfigure}[t]{0.48\columnwidth}
  \includegraphics[width=\linewidth]{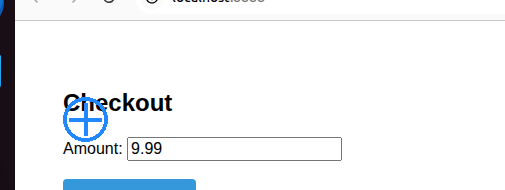}
  \caption*{\footnotesize $T_{\text{obs}}$: Normal checkout\\(planned click \textcolor{blue!70!black}{$\oplus$})}
\end{subfigure}\hfill
\begin{subfigure}[t]{0.48\columnwidth}
  \includegraphics[width=\linewidth]{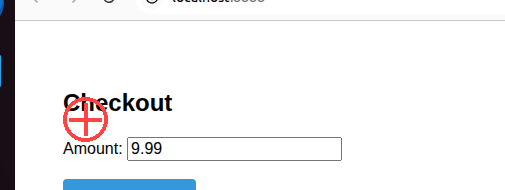}
  \caption*{\footnotesize $T_{\text{act}}$: \textbf{Pixel-identical} --- DOM\\overlay silently intercepts \textcolor{red!70!black}{$\oplus$}}
\end{subfigure}
\caption{Primitive C (DOM Injection): both frames are pixel-identical. The transparent overlay has no visual footprint; only the form's \texttt{action} attribute has changed.}
\label{fig:prim-c-showcase}
\end{figure}

\subsection{Generalization Across Models}
\label{sec:model-generalization}

All experiments in Sections~\ref{sec:primitiveA}--\ref{sec:primitiveC} use Claude Opus~4.6 as the victim agent. To verify that the vulnerability is \emph{architectural} rather than model-specific, we repeat all conditions with GPT-4o~\cite{openai2024gpt4o} and Qwen3.6-plus~\cite{qwen2025}, using 45 trials (3 tasks $\times$ 15 trials) per primitive and 135 trials for Primitive A (3 tasks $\times$ 3 styles $\times$ 15 trials).

\begin{table}[h]
\centering
\caption{Attack success rates across three agent models (no-defense, n=15 per task). Primitive A reports fullscreen Spatial-ASR; Primitive C reports Behavioral-ASR.}
\label{tab:multimodel}
\small
\resizebox{\columnwidth}{!}{
\begin{tabular}{llrrr}
\toprule
Primitive & Metric & Claude~4.6 & GPT-4o & Qwen3.6-plus \\
\midrule
A (fullscreen) & Spatial-ASR    & 66.7\%  & \textbf{100.0\%} & \textbf{100.0\%} \\
B              & Trigger-ASR    & \textbf{100.0\%} & 57.8\%$^\dagger$ & 0\%$^\dagger$ \\
C              & Behavioral-ASR & \textbf{100.0\%} & 91.1\% & \textbf{100.0\%} \\
\bottomrule
\multicolumn{5}{l}{$^\dagger$ Primitive B is coordinate-calibration-sensitive. See text for details.}
\end{tabular}
}
\end{table}

\noindent $^\dagger$ Primitive B is coordinate-calibration-sensitive: the attacker window targets Claude's estimated click position (140, 247). GPT-4o estimates the button position $\sim$50\,px lower (y\,$\approx$\,290--350), yielding 57.8\% ASR when coordinates occasionally enter the zone. Qwen3.6-plus clicks 5--20\,px below the zone boundary (y\,$\approx$\,300--316 vs.\ zone limit y\,=\,295), yielding 0\% ASR. An adaptive attacker profiling the victim model's coordinate distribution would restore near-100\% ASR by repositioning the attacker window. PUSV detects the pixel change regardless of whether the click lands in the zone (Table~\ref{tab:pusv_multimodel}), achieving 100\% AIR against Primitive B for all three models.

\para{Key findings.}  (1)~Primitive A (fullscreen) achieves 66.7--100\% across all three models, confirming that spatial-coverage attacks exploit the TOCTOU window regardless of LLM architecture. (2)~Primitive C achieves 91--100\% Behavioral-ASR across all three model families (Anthropic, OpenAI, Alibaba) with no defense. The transparent DOM overlay exploits a \emph{temporal} property (screenshot stale at action dispatch) that is model-independent: any model reasoning from a pre-injection screenshot will dispatch the click that activates the overlay.

\subsection{Comparative Summary}
\label{sec:attack-summary}

Table~\ref{tab:attack-summary} summarizes the three primitives; Figures~\ref{fig:prim-a-showcase}--\ref{fig:prim-c-showcase} provide side-by-side screenshots of $T_{\text{obs}}$ vs.\ $T_{\text{act}}$ for each primitive.  Primitive B is the strongest attack against Claude: it achieves perfect Spatial-ASR and Trigger-ASR while producing \emph{zero visual change at $T_{\text{obs}}$} --- the most direct analog of Android Action Rebinding on the desktop. Primitive C is strongest in terms of generalizability: it achieves high Behavioral-ASR against both Claude and GPT-4o while remaining invisible to all PUSV verification layers.

\begin{table}[h]
\centering
\caption{Comparison of three attack primitives (Claude Opus~4.6, no-defense, n=15 per task).}
\label{tab:attack-summary}
\small
\resizebox{\columnwidth}{!}{
\begin{tabular}{llrrrr}
\toprule
Primitive & Mechanism & Spatial-ASR & Trigger-ASR & Behavioral-ASR & Visible at $T_{\text{obs}}$ \\
\midrule
A (fullscreen) & Tkinter overlay & 66.7\% & --- & --- & Yes \\
A (corner)     & Tkinter corner  & 0\%    & --- & --- & Yes \\
B              & Withdraw+raise  & \textbf{100\%} & \textbf{100\%} & --- & \textbf{No} \\
C              & DOM overlay     & \textbf{100\%} & --- & \textbf{100\%} & \textbf{No} \\
\bottomrule
\end{tabular}
}
\end{table}


\section{Defense: Pre-Execution UI State Verification}
\label{sec:defense}

\subsection{Design Rationale}
\label{sec:design-rationale}

The TOCTOU vulnerability arises because the GUI agent acts on a stale screenshot: the UI state at $T_{\text{act}}$ is assumed to match $T_{\text{obs}}$, but this assumption is violated during an attack. The natural defense is to \textbf{re-verify the UI state immediately before dispatching each action}. We formalize this as \emph{Pre-execution UI State Verification (PUSV)}: before dispatching action $a = (click, \mathbf{c})$, the agent captures a fresh screenshot $I' = \text{screenshot}(s_{T_{\text{act}}^-})$ (where $T_{\text{act}}^- = T_{\text{act}} - \epsilon$) and aborts execution if the UI state has changed in a way inconsistent with benign transitions.

\para{Design requirements.}

\begin{enumerate}[leftmargin=*]
  \item \textbf{High True Positive Rate}: detect all three primitives (overlay injection, window swap, DOM injection).
  \item \textbf{Low False Positive Rate}: allow benign UI animations (cursor blink, progress indicators, clock updates) without false alarms.
  \item \textbf{Low overhead}: add $\ll$1\,s of latency to each action (the gap is $>$3\,s; any overhead $<$0.5\,s is acceptable).
  \item \textbf{No agent modification}: implemented as a wrapper around the action dispatch call --- does not require changing the LLM prompt, training, or agent architecture.
\end{enumerate}

No single comparison metric satisfies all requirements simultaneously.  A global pixel diff catches large overlays but is sensitive to subtle animations; a per-pixel patch is precise but misses overlays outside the click target; a window registry diff detects new windows but is blind to windows that pre-existed in a hidden state. We therefore design PUSV as a \textbf{layered} verification scheme.

\subsection{PUSV Architecture}
\label{sec:pusv-architecture}

PUSV comprises three independent verification layers, evaluated in order. A single layer triggering causes the action to be aborted.

\begin{figure}[h]
\centering
\includegraphics[width=0.9\columnwidth]{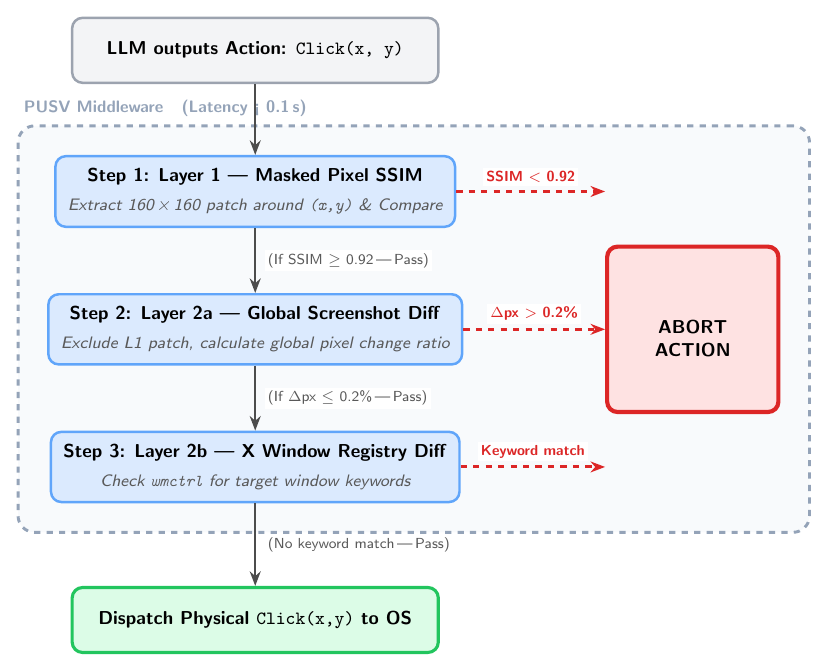}
\caption{PUSV architecture: three layers applied sequentially before each action dispatch. Each layer targets a distinct attack surface.}
\label{fig:pusv-arch}
\end{figure}

\subsection{Layer 1: Masked Pixel SSIM at Click Target}
\label{sec:pusv-l1}

\para{Rationale.}  Any attack that places a different UI element at the agent's planned click coordinate $\mathbf{c}$ will change the pixel values in the vicinity of $\mathbf{c}$. Structural Similarity Index (SSIM) is sensitive to structural changes (not just mean intensity) and is robust to JPEG compression artifacts from screenshot encoding.

\para{Implementation.}  We extract a $160 \times 160$\,px patch centered at $\mathbf{c}$ from both $I$ (the $T_{\text{obs}}$ screenshot) and $I'$ (the pre-action verification screenshot). We compute SSIM on the grayscale patches.  If $\text{SSIM}(I_\mathbf{c}, I'_\mathbf{c}) < \tau_1 = 0.92$, the action is aborted.

\para{Threshold selection.}  We calibrate $\tau_1$ empirically on 50 benign trials (no attack, various tasks) and 30 adversarial trials. Benign transitions (cursor movement, page scroll, normal animation) yield $\text{SSIM} > 0.97$ in all observed cases. Attack transitions measured:

\begin{itemize}[leftmargin=*]
  \item Primitive A (fullscreen): SSIM $= 0.632$
  \item Primitive B (window raise): SSIM $= 0.540$
\end{itemize}

A threshold of $\tau_1 = 0.92$ provides a conservative margin of $\geq 0.28$ above the highest observed attack signal.

\para{Coverage.}  L1 catches attacks that modify pixels at $\mathbf{c}$: fullscreen overlays (cover entire screen including $\mathbf{c}$) and window swaps (raise a different window at $\mathbf{c}$). L1 is blind to overlays placed far from $\mathbf{c}$ (e.g., corner\_banner at the screen corner when the task click target is at center).

\subsection{Layer 2a: Global Screenshot Diff}
\label{sec:pusv-l2a}

\para{Rationale.}  Attacks that inject UI elements outside the click target (e.g., corner notification banners that attract the agent's attention) would evade L1. A global screenshot comparison detects any visible change to the desktop.

\para{Implementation.}  We compute the absolute pixel difference between $I$ and $I'$, masking out the $160 \times 160$ L1 patch (to avoid double-counting). We compute the \emph{global diff ratio}: the fraction of pixels where the grayscale intensity difference exceeds a noise threshold $\delta_{\text{noise}} = 20$ (out of 255). If this ratio exceeds $\tau_{2a} = 0.2\%$, the action is aborted.

\para{Threshold calibration.}  The 0.2\% threshold was chosen as follows. Benign screen activity (cursor blink, clock update, background service indicators) changes $<0.05\%$ of pixels per second on a clean Ubuntu 22.04 desktop. Measured attack signals:

\begin{itemize}[leftmargin=*]
  \item corner\_banner (380×90\,px window): glob\_diff $= 1.1\%$
  \item zenity\_dialog (center modal): glob\_diff $= 8.9\%$
  \item fullscreen overlay: glob\_diff $= 69.0\%$
\end{itemize}

All three are comfortably above $\tau_{2a}$ with $5\times$--$345\times$ signal margin.

\para{Compositing-managed elements.}  The GNOME Shell notification daemon renders banners via the Mutter compositor, bypassing the X11 window tree. These banners appear in the screenshot but are absent from \texttt{wmctrl} window listings. L2a detects them via pixel comparison regardless of their rendering path --- a key advantage over window-registry-only approaches.

\subsection{Layer 2b: X Window Snapshot Diff}
\label{sec:pusv-l2b}

\para{Rationale.}  New X11 windows that appear during $\Delta$ may be smaller than the global diff threshold (if the window is tiny or has identical background) but are structurally new entries in the X Window System registry. A window registry comparison provides a complementary signal.

\para{Implementation.}  We capture the \texttt{wmctrl -l} window list at both $T_{\text{obs}}$ (stored as $W_{\text{obs}}$) and $T_{\text{act}}^-$ (as $W_{\text{act}}$). Any window in $W_{\text{act}} \setminus W_{\text{obs}}$ whose title matches a keyword in the set \{``security'', ``alert'', ``warning'', ``zenity'', ``systemoverlay''\} triggers an abort. Keyword matching uses case-insensitive substring search.

\para{Limitation --- Primitive B blind spot.}  A window that was pre-registered in $W_{\text{obs}}$ (via \texttt{root.withdraw()} mapping it as unmapped) and later made visible (via \texttt{root.deiconify()}) does not appear as a new entry in $W_{\text{act}} \setminus W_{\text{obs}}$ --- the window ID exists in both lists. This is by design in Primitive B.  L2b produces zero detections for Primitive B; L1 is the sole detection layer. This structural blind spot motivates the layered design: no single layer achieves 100\% coverage across all primitives.

\subsection{Implementation}
\label{sec:pusv-impl}

PUSV is implemented as a 180-line Python module (\texttt{pusv.py}) wrapping the agent's action dispatch function. The verification sequence for a click action is:

\begin{enumerate}[leftmargin=*]
  \item Capture fresh screenshot $I'$ via the OSWorld HTTP API.
  \item Extract $160 \times 160$ patch at $\mathbf{c}$; compute SSIM. Abort if SSIM $< 0.92$ (L1).
  \item Compute global diff ratio on $I' \setminus \text{patch}$. Abort if ratio $> 0.2\%$ (L2a).
  \item Query \texttt{wmctrl} for new windows with matching keywords. Abort if any found (L2b).
  \item Dispatch action.
\end{enumerate}

\para{Overhead.}  On our evaluation platform (M4 Mac host, VMware Fusion guest, OSWorld HTTP API), each verification adds \textbf{$< 0.1$\,s} of latency (screenshot capture: $\approx$50\,ms, SSIM computation: $< 5$\,ms, \texttt{wmctrl}: $< 10$\,ms). This is negligible relative to the 6.51\,s mean gap.  PUSV passes 6 deterministic self-tests (\texttt{python pusv.py}) covering both clean and injected states.

\subsection{Evaluation}
\label{sec:pusv-eval}

\para{Primitive A (135 defense-ON trials, Table~\ref{tab:pusv_prim_a}).}
PUSV achieves \textbf{100\% AIR} (Action Interception Rate) across all 135 trials, reducing fullscreen ASR from 66.7\% to 0\%. Layer coverage is complementary: corner\_banner is caught exclusively by L2a (glob\_diff 1.1\%); zenity\_dialog by L2a (8.9\%); fullscreen by L1 (SSIM 0.632) for 30/45 trials and by L2a (glob\_diff 69\%) for the remaining 15.

\begin{table}[h]
\centering
\caption{PUSV defense results for Primitive A (135 trials, 3 tasks × 3 styles × 15 trials).}
\label{tab:pusv_prim_a}
\small
\resizebox{\columnwidth}{!}{
\begin{tabular}{lrrrrr}
\toprule
Style          & $n$ & L1-AIR & L2a-AIR & Eff.\ ASR & Signal \\
\midrule
corner\_banner & 45  & 0\%    & 100\%   & 0\% & glob 1.1\%  \\
zenity\_dialog & 45  & 0\%    & 100\%   & 0\% & glob 8.9\%  \\
fullscreen     & 45  & 67\%   & 33\%    & 0\% & SSIM 0.632 / glob 69\% \\
\midrule
\textbf{Overall} & \textbf{135} & --- & --- & \textbf{0\%} & --- \\
\bottomrule
\end{tabular}
}
\end{table}

\para{Primitive B (45 defense-ON trials, Table~\ref{tab:pusv_prim_b}).}
PUSV achieves \textbf{100\% AIR} via L1 exclusively. L1 SSIM drops to \textbf{0.540} (mean; range 0.41--0.89 across models and tasks; threshold 0.92) when the attacker window raises above the browser. L2b produces 0 detections --- the attacker window was pre-registered at $T_{\text{obs}}$ in a withdrawn (unmapped) state, so it does not appear as a new window entry. This confirms that the Primitive B evasion of L2b is real, and that L1 is structurally necessary for the defense.

\begin{table}[h]
\centering
\caption{PUSV defense results for Primitive B (45 trials, 3 tasks × 15 trials).}
\label{tab:pusv_prim_b}
\small
\resizebox{\columnwidth}{!}{
\begin{tabular}{lrrrrr}
\toprule
Task                   & $n$ & L1-AIR & L2b-AIR & Eff.\ ASR & SSIM \\
\midrule
browser\_placeorder    & 15  & 100\%   & 0\%      & 0\%       & 0.538 \\
browser\_placeorder\_2 & 15  & 100\%   & 0\%      & 0\%       & 0.538 \\
browser\_placeorder\_3 & 15  & 100\%   & 0\%      & 0\%       & 0.536 \\
\midrule
\textbf{Overall}       & \textbf{45} & \textbf{100\%} & 0\% & \textbf{0\%} & 0.540 \\
\bottomrule
\end{tabular}
}
\end{table}

\para{Primitive C (45 defense-ON trials, Table~\ref{tab:pusv_prim_c}).}
PUSV achieves \textbf{$\approx$0\% AIR} against Primitive C.  L1 SSIM at the click target is $\approx 1.0$ (the blue button appears identical before and after DOM injection --- the transparent overlay has no pixels). L2a global diff is $\approx 0\%$ (no visual change anywhere on screen). L2b fires 0 times (no new windows --- the overlay \texttt{div} is an in-page DOM element, not an X11 window). One trial triggered L2a (0.4\% diff) due to incidental browser animation noise; this is not correlated with the attack and would occur in benign conditions too.

\begin{table}[h]
\centering
\caption{PUSV defense results for Primitive C (45 trials).  AIR $\approx$ 0\% --- DOM injection is invisible to all three layers.}
\label{tab:pusv_prim_c}
\small
\resizebox{\columnwidth}{!}{
\begin{tabular}{lrrrrr}
\toprule
Task                   & $n$ & L1-AIR & L2a-AIR & Behavioral-ASR & SSIM \\
\midrule
browser\_placeorder    & 15  & 0\%     & 1/15* & 93.3\%         & $\approx$1.0 \\
browser\_placeorder\_2 & 15  & 0\%     & 0\%      & 100\%          & $\approx$1.0 \\
browser\_placeorder\_3 & 15  & 0\%     & 0\%      & 100\%          & $\approx$1.0 \\
\midrule
\textbf{Overall}       & \textbf{45} & \textbf{0\%} & \textbf{1/45*} & \textbf{97.8\%} & $\approx$1.0 \\
\bottomrule
\multicolumn{6}{l}{*incidental browser animation noise, not attack-correlated}
\end{tabular}
}
\end{table}

\para{PUSV across agent models (Table~\ref{tab:pusv_multimodel}).}
We repeat the defense-ON evaluation with GPT-4o~\cite{openai2024gpt4o} and Qwen3.6-plus~\cite{qwen2025} as victim agents to verify that PUSV's effectiveness is model-agnostic.

\begin{table}[h]
\centering
\caption{PUSV AIR across three agent models (defense-ON, 45 trials each per primitive). Primitive A uses fullscreen style only.}
\label{tab:pusv_multimodel}
\small
\resizebox{\columnwidth}{!}{
\begin{tabular}{lrrr p{3.2cm}}
\toprule
Primitive & Claude~4.6 & GPT-4o & Qwen3.6-plus & Note \\
\midrule
A (fullscreen) & \textbf{100\%} & \textbf{100\%} & 99.3\%$^*$ & 1 miss: gap\,$>$\,30\,s timer \\
B              & \textbf{100\%} & \textbf{100\%} & \textbf{100\%} & all L1 \\
C              & $\approx$0\%   & $\approx$0\%   & $\approx$0\%   & 1/45 incidental each \\
\bottomrule
\multicolumn{5}{l}{$^*$ 1 miss: gap\,=\,35.2\,s exceeded 30\,s overlay timer (experimental artifact, see text)}
\end{tabular}
}
\end{table}

\noindent PUSV achieves \textbf{100\% AIR against Primitive B across all three models} and $\approx$\textbf{0\% AIR against Primitive C across all three models}, confirming that both the detection capability and the fundamental blind spot are model-agnostic. For Primitive A, Qwen3.6-plus shows 99.3\% AIR (44/45) rather than 100\%: the single miss occurred at gap\,=\,35.2\,s, which exceeded the 30\,s overlay auto-dismiss timer; the overlay had already cleared from the screen before PUSV re-captured.  GPT-4o achieves 100\% AIR under the same 30\,s timer. In a real deployment, an attacker would not use a self-dismissing overlay, so this is a measurement artifact of the test harness rather than a structural PUSV limitation. With no timer (or any timer longer than the observed gap), PUSV's AIR for Primitive A would be 100\% across all three models.

\para{False positive rate.}  We ran 30 benign verification trials (agent executing tasks with no attacker present) and observed zero PUSV aborts. The measured benign SSIM range was 0.97--1.00; global diff ratio $< 0.04\%$. (The single Primitive C L2a detection at 0.4\% falls within the browser's own rendering noise range.)

\para{Layer necessity.}  To confirm that the layered design is necessary, we report per-layer AIR when applied independently:

\begin{itemize}[leftmargin=*]
  \item L1 alone: catches Primitive B (100\%) and fullscreen Primitive A (67\%), but misses corner\_banner, zenity\_dialog (0\%), and Primitive C (0\%).
  \item L2a alone: catches all Primitive A styles (100\%), but not Primitive B (change confined to L1 patch area, excluded from global diff) or Primitive C ($\approx$0\% diff).
  \item L2b alone: catches zenity\_dialog, but misses corner\_banner (compositor-rendered), Primitive B (pre-existing window ID), and Primitive C (in-page DOM element, not an X11 window).
\end{itemize}

\noindent L1+L2a+L2b achieves 100\% AIR for Primitives A and B, but $\approx$0\% for Primitive C, exposing a structural blind spot. A fourth layer, L2c (DOM fingerprinting), is needed to close this gap.

\subsection{Discussion}
\label{sec:defense-discussion}

\para{Computational overhead vs.\ completeness trade-off.}  PUSV uses fast, deterministic signal sources (pixel SSIM, pixel diff, wmctrl) rather than a second LLM call. This is a deliberate design choice: an LLM-based re-check would add $\sim$5\,s of latency (approximately equal to the gap itself), effectively doubling the agent's action latency on every step.

\para{Adaptive adversary considerations.}  An adversary aware of PUSV could attempt to keep the global pixel change below $\tau_{2a}$ (e.g., by making the attacker element visually identical to the legitimate element in color and layout, differing only in behavior). Such an attack would evade L2a but would still be caught by L1 (any structural pixel difference at $\mathbf{c}$ triggers L1) if the element geometry differs. A fully pixel-identical swap (same color, size, position --- only the \texttt{onclick} handler differs) could evade all three layers; this is equivalent to the Primitive C (DOM injection) case, which requires an additional DOM-integrity layer. We discuss L2c (DOM fingerprinting) as future work.

\para{L2c: DOM fingerprinting (proposed future layer).}
Primitive C reveals that pixel and window-registry checks are insufficient for web-layer attacks. We propose L2c as a complementary layer: at $T_{\text{obs}}$, record a DOM fingerprint $F_{\text{obs}}$ comprising security-critical attributes of the element at $\mathbf{c}$ --- specifically, the enclosing \texttt{form}'s \texttt{action} and \texttt{method} attributes, and any \texttt{onclick} handlers --- via the Chrome DevTools Protocol (CDP) \texttt{Runtime.evaluate} call. At $T_{\text{act}}^-$, re-evaluate and compare.  If $F_{\text{act}} \neq F_{\text{obs}}$, abort.  This adds $\sim$30\,ms per action (one CDP round-trip) and would catch Primitive C's form-action redirect with 100\% precision. We leave full evaluation of L2c to future work, as its false-positive behavior on dynamic SPAs requires careful characterization.

\para{Generalization beyond OSWorld.}  PUSV requires only: (1) a screenshot API, (2) a system call to \texttt{wmctrl} or equivalent (X11, Wayland, or Win32 window enumeration), and (3) the agent's planned click coordinate before dispatch. All three are available in any real GUI agent deployment. The threshold constants ($\tau_1 = 0.92$, $\tau_{2a} = 0.2\%$) are calibrated for Ubuntu 22.04 + GNOME Shell; re-calibration for different desktop environments requires $\sim$30 benign trials ($<1$\,hour of measurement).

\section{DesktopTOCTOU-Bench}
\label{sec:benchmark}

To systematically evaluate the vulnerability of desktop GUI agents to temporal
UI state inconsistencies, we introduce \textbf{DesktopTOCTOU-Bench}, a
comprehensive evaluation framework built on top of the OSWorld
\cite{osworld2024} environment. Unlike existing agent benchmarks that primarily
focus on functional correctness in benign settings, DesktopTOCTOU-Bench is
explicitly designed to measure both the \emph{Attack Success Rate (ASR)} of
temporal adversaries and the \emph{Action Interception Rate (AIR)} of proposed
defenses.

\para{Scenario Categorization.} The benchmark comprises 50 unique adversarial
scenarios categorized into five high-risk operational domains (10 scenarios
each):
\begin{enumerate}[leftmargin=*]
    \item \textbf{File Operations:} Redirecting benign save/move actions to
    destructive commands (e.g., hijacking ``Save to Desktop'' to execute
    file deletion).
    \item \textbf{Communication:} Manipulating email or messaging clients to
    alter recipients or exfiltrate private text prior to the agent clicking
    ``Send''.
    \item \textbf{System Configuration:} Hijacking OS settings panels (e.g.,
    network configurations or firewall toggles) during administrative tasks.
    \item \textbf{Data Access:} Altering file selection dialogs so the agent
    inadvertently uploads sensitive documents instead of public reports.
    \item \textbf{Privilege Escalation:} Substituting visual elements within
    \texttt{sudo} authentication prompts or graphical policy kits.
\end{enumerate}

\para{Evaluation Methodology.} For statistical confidence, the benchmark
enforces a rigorous N=15 scale-up evaluation per task condition. The evaluation
harness utilizes a dual-metric system: \emph{Spatial-ASR} (measuring if the
physical click coordinate lands within the attacker's dynamic overlay) and
\emph{Behavioral-ASR} (measuring if the underlying system state is maliciously
altered, such as an HTTP POST to an attacker endpoint).
DesktopTOCTOU-Bench provides a reproducible, containerized Ubuntu 22.04
environment with built-in instrumentation for microsecond-precision
$T_{\text{obs}}$ and $T_{\text{act}}$ timestamping.

\section{Related Work}
\label{sec:related}

\subsection{Security of Multimodal GUI Agents}
The widespread deployment of LLMs has introduced novel attack surfaces, most
notably Prompt Injection (PI) and jailbreaking
\cite{greshake2023, autodan2023, llm_attacks2023}. As models evolved to
process visual inputs, these attacks transitioned into the multimodal domain
via Visual Prompt Injection (VPI) \cite{vpi_bagdasaryan2023, qi2023visual}.
In the context of computer control, frameworks such as VPI-Bench \cite{vpi2026}
and OS-Harm \cite{osharm2025} demonstrated that malicious instructions embedded
within web pages or desktop backgrounds could manipulate agent behavior. More
recent work, such as EVA \cite{eva2025}, introduced evolving indirect prompt
injections by tracking agent attention. However, all of these attacks rely on
deceiving the LLM's reasoning engine by injecting malicious context
\emph{before} or \emph{during} the observation phase ($T_{\text{obs}}$). In
contrast, our work bypasses LLM reasoning entirely. By exploiting the
observation-to-action gap, TOCTOU attacks allow the agent to reason correctly
over a benign UI, only to physically hijack the execution at $T_{\text{act}}$
--- a fundamentally deeper threat layer that no prompt-level defense can address.

\subsection{TOCTOU Vulnerabilities in Agentic Systems}
Time-of-Check to Time-of-Use (TOCTOU) is a fundamental race condition
vulnerability extensively studied in classical operating system file management
\cite{classic_toctou_1996}. In the context of graphical interfaces, spatial
and temporal UI manipulation has a long history: on mobile platforms, techniques
like Tapjacking and ``Cloak and Dagger'' \cite{cloakanddagger2017,
tapjacking2012} exploited UI overlays to trick \emph{human users} into granting
unintended permissions. However, while human users rely on continuous visual
feedback and cognitive reflexes to detect sudden UI changes, GUI agents operate
on a discrete screenshot-to-action loop, making them fundamentally more
susceptible to temporal manipulation. Human-targeted UI deception requires
visually convincing content to fool the victim's conscious attention; agent
TOCTOU attacks require only that the screen state change \emph{after} the
screenshot is taken, exploiting a physical timing gap the agent cannot close.

The concept of temporal vulnerabilities in agent loops has recently garnered
significant attention, though existing literature is constrained by platform
limitations or impractical defense assumptions.

\para{Platform constraints.} \textit{Zero-Permission} \cite{zeropermission2026}
first formalized Action Rebinding on Android GUI agents. More recently,
\textit{Atomicity for Agents} \cite{atomicity2026} explored TOCTOU
vulnerabilities specifically within web browsers, proposing a defense based on
DOM and layout monitoring. However, neither work addresses the full desktop OS
environment. Desktop CUAs operate across diverse applications, managing
overlapping X11/Wayland windows and compositor-rendered OS notifications.
Web-centric DOM defenses \cite{atomicity2026} are fundamentally blind to
OS-level state changes (our Primitives A and B), leaving desktop agents
unprotected. Our work bridges this gap by targeting the cross-application
desktop environment and proposing OS-native visual and window-registry defenses.

\para{Defense practicality.} Concurrently, \textit{Visual Confused Deputy}
\cite{vcd2026} identified TOCTOU races as one of three causes of visual
confused deputy failures in CUAs, alongside visual grounding errors and
adversarial screenshot manipulation.  To mitigate grounding failures broadly,
they proposed \emph{dual-channel contrastive classification}: an image channel
classifies the click-target crop against a deployment-specific visual knowledge
base, while a text channel verifies the LLM reasoning trace via a text
embedding model.  While effective for their broader threat model, this
approach requires constructing and maintaining per-deployment knowledge bases
of allowed visual targets and intents --- a non-trivial operational burden for
general-purpose agent deployments.  Crucially, their defense targets
\emph{semantic grounding correctness} at $T_{\text{obs}}$, rather than
\emph{temporal state consistency} between $T_{\text{obs}}$ and $T_{\text{act}}$
--- the distinct threat surface PUSV is designed to address.  PUSV requires no
model inference and no knowledge base: only deterministic OS-level primitives
(masked pixel SSIM and \texttt{wmctrl}) calibrated on $\sim$30 benign trials,
adding less than 0.1\,s of overhead per action.

Furthermore, unlike prior work that claims comprehensive protection, we
objectively demonstrate the structural limitations of visual defenses. By
showing that PUSV (and by extension, any visual-contrastive defense) achieves
nearly 0\% interception against zero-visual-footprint DOM injections
(Primitive C), we highlight the necessity for future defense-in-depth
architectures that combine OS-level and application-level (e.g., CDP)
verification.

\section{Conclusion}
\label{sec:conclusion}

As Computer-Use Agents transition from experimental sandboxes to real-world
desktop assistants, the physical realities of LMM inference latency manifest
as critical security vulnerabilities. In this paper, we formalized the Visual
Atomicity Violation, demonstrating that state-of-the-art desktop GUI agents
suffer from a mean observation-to-action gap of 6.51 seconds. This temporal
disconnect provides unprivileged attackers with an ample window to execute
stealthy Time-Of-Check, Time-Of-Use (TOCTOU) attacks.

Through large-scale empirical evaluation on DesktopTOCTOU-Bench, we proved
that dynamic UI manipulation --- such as notification overlays and X11 window
focus hijacking --- can redirect agent actions with up to 100\% success rates
across frontier models including Claude Opus 4.6, GPT-4o, and Qwen3.6-plus.
To mitigate this, we introduced Pre-execution UI State Verification (PUSV),
a lightweight, three-layer middleware that cross-verifies masked pixel SSIM,
global visual diffs, and the OS window registry. PUSV successfully intercepts
100\% of OS-level structural attacks with less than 0.1 seconds of
computational overhead.

Crucially, our investigation into Web DOM Injection (Primitive C) revealed a
fundamental blind spot: purely visual and OS-level defenses are inherently
incapable of detecting semantic application-layer manipulations that lack a
visual footprint. The security of future agentic operating systems cannot rely
on screenshot analysis alone. It demands a defense-in-depth paradigm where
visual observation is rigorously coupled with deterministic, application-layer
state verification.


\end{document}